# Machine Learning-Assisted Thermoelectric Cooling for On-Demand Multi-Hotspot Thermal Management

Jiajian Luo[1] and Jaeho Lee[1,a)]

[1]Department of Mechanical and Aerospace Engineering, University of California Irvine, Irvine, CA, 92697, USA

Thermoelectric coolers (TECs) offer a promising solution for direct cooling of local hotspots and active thermal management in advanced electronic systems. However, TECs present significant trade-offs among spatial cooling, heating and power consumption. The optimization of TECs requires extensive simulations, which are impractical for managing actual systems with multiple hotspots under spatial and temporal variations. In this study, we present a novel machine learning-assisted optimization algorithm for thermoelectric coolers that can achieve global optimal temperature by individually controlling TEC units based on real-time multi-hotspot conditions across the entire domain. We train a convolutional neural network (CNN) with a combination of the Inception module and multi-task learning (MTL) approach to comprehend the coupled thermal-electrical physics underlying the system and attain accurate predictions for both temperature and power consumption with and without TECs. Due to the intricate interaction among passive thermal gradient, Peltier effect and Joule effect, a local optimal TEC control experiences spatial temperature trade-off which may not lead to a global optimal solution. To address this issue, we develop a backtracking-based optimization algorithm using the machine learning model to iterate all possible TEC assignments for attaining global optimal solutions. For any m × n matrix with $N_{HS}$ hotspots (n, m ≤ 10, 1 ≤ $N_{HS}$ ≤ 20), our algorithm is capable of providing 52.4% peak temperature reduction and its corresponding TEC array control within an average of 1.64 seconds while iterating through tens of temperature predictions behind-the-scenes. This represents a speed increase of over three orders of magnitude compared to traditional FEM strategies which take approximately 27 minutes.

**I. INTRODUCTION**

Despite great advancements in semiconductor technology beyond the sub-3nm node[1], most thermal management techniques nowadays are limited to the macroscale operation. The trend towards device miniaturization and the rapid emergence of System-on-Chip (SoC) inevitably complicate the thermal behavior within microelectronic devices[2,3]. Specifically, multiple on-chip hotspots exhibit spatial and temporal changes due to workload variations, environmental fluctuations, device defects and aging, which can occur among modules[4,5], cores (processors)[6,7] and transistors[8]. The complexity of the hotspot behavior presents unprecedented challenges for conventional thermal management methods which only rely on uniform control, necessitating a more efficient, sophisticated, and intelligent approach capable of on-demand thermal management to ensure optimal functionality and longevity of microelectronic devices[9].

Among various active cooling techniques, thermoelectric coolers (TECs) offer distinctive local cooling capability as well as several other advantages[10–12], making them a promising solution to hotspot thermal management. In recent years, there have been emerging designs utilizing single TECs[13–16] and TEC arrays[17–19] for on-chip hotspot cooling in microelectronic devices. Revolutionary materials, including nanostructured Si[20–22], self-hygroscopic hydrogel[23] and flexible inorganics[24,25],

---

a) Correspondence author. Email: jaeholee@uci.edu.

are extensively studied to improve TEC cooling performance. However, TEC cooling exhibits significant trade-offs in spatial temperature and power consumption, and its performance relies on multiple variables including TEC voltages and hotspot conditions[12,26]. The high non-linearity in TEC behavior requires multiple solutions for optimization, which brings expensive computational cost to conventional finite element method (FEM) simulations. Furthermore, in actual applications where multiple hotspots undergo spatial and temporal evolution, the traditional techniques become more challenging and even impossible to realize a real-time optimal TEC control.

The thriving field of machine learning offers a powerful tool for thermoelectric research by providing neural network models that greatly expedite the process of thermoelectric material selection[27], TEC design[28,29] and optimization[30,31]. However, these models primarily focus on the analysis of back-end designs by considering an individual, isolated TEC device[28,29], or using over-simplified optimization logics such as linear control[30] and uniform control[31]. Apparently, there is still vacancy and urgent need for a more comprehensive model that can comprehend the coupled thermal-electrical physics in TECs while predicting their spatial interplays with multiple hotspots undergoing dynamic evolution across the entire domain. Ultimately, this model should be capable of providing responsive TEC control and the corresponding power consumption over the entire domain to achieve real-time global optimal temperature.

In this study, we present a machine learning-assisted optimization algorithm for TECs that fulfills the aforementioned on-demand thermal management. We utilize our previous holey silicon-based TEC array with independent TEC control[32] as an illustrative example for conducting the analysis. We develop a convolutional neural network (CNN) with the Inception module and multi-task learning (MTL) approach to perceive the spatial correlation of TECs and hotspots, thereby accurately predicting temperature and power consumption by comprehending the thermal-electrical physics underlying the system. During the TEC optimization process, the major challenge lies in the intricate thermal-electrical interaction among multiple hotspots and TECs, since a local optimal TEC control may not lead to a global optimal solution due to temperature redistribution. Therefore, we develop a backtracking-based optimization algorithm that efficiently explores all potential TEC assignments in order to obtain the global optimal temperature based on real-time hotspot conditions. Note that this methodology can be applied to general TEC / TEC array designs with a wide range of thermoelectric materials (e.g., $Bi_2Te_3$/$Sb_2Te_3$), configurations (e.g., lateral- and vertical-oriented TECs) and device scales (e.g., module-scale and transistor-scale). Consequently, this approach can hopefully provide efficient TEC/ TEC array control logics for the future TEC-incorporated electronic systems.

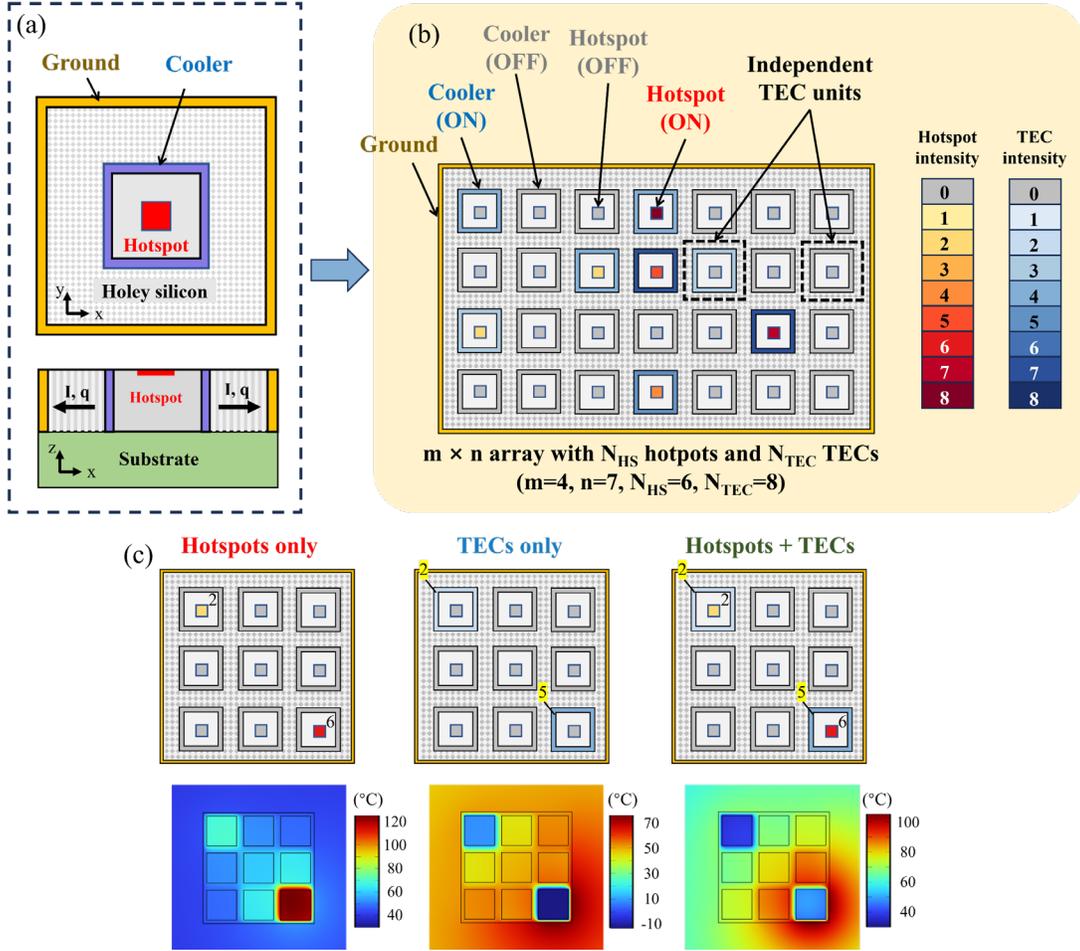

FIG. 1. Holey silicon-based lateral TEC and its array model. (a) The schematic of a single TEC. (b) The schematic of an arbitrary m × n TEC array with $N_{HS}$ assigned hotspots and $N_{TEC}$ assigned TECs (1 ≤ m, n ≤ 10; 0 ≤ $N_{HS}$, $N_{TEC}$ ≤ min[m × n, 20]). The intensities of hotspots and TECs are in 9 discrete levels. (c) FEM modelling examples of a 3 × 3 hotspot-TEC array with three scenarios: hotspots only, TECs only and hotspots + TECs.

## II. TEC MODELLING

To demonstrate our study, we choose our previous theoretical designs[32], the holey silicon-based singe TEC (Fig. 1a) and its scaling array (Fig.1b) as our TEC model. The model features a lateral orientation of the TEC components (i.e., Peltier electrodes and holey silicon region) along with the central hotspot. Here, holey silicon is the thermoelectric material due to its compatibility with microfabrication processes, and because the introduction of vertical nanoholes results in substantial decrease of in-plane thermal conductivity due to phonon boundary scattering, meanwhile retaining excellent electrical properties (electrical conductivity and Seebeck coefficient) from p-type silicon[15,21,33–35]. When positive voltage is applied to the cooler, lateral heat redistribution occurs which provides active heat removal from the hotspot to the in-plane surroundings. Compared to a single TEC, the TEC array as shown in Fig. 1b has multiple coolers enclosed by a single ground that allow for independent TEC control. Based on the existing hotspot conditions, different coolers can accept different input voltages to achieve on-demand thermal management.

In general, the TEC modelling involves thermal-electrical physics which includes passive heat diffusion, thermoelectric effect and Joule effect. In steady state, the governing equations can be written as:

$$\nabla \cdot \boldsymbol{q} = Q_e \tag{1}$$

$$Q_e = \boldsymbol{J} \cdot \boldsymbol{E} \tag{2}$$

$$\boldsymbol{q} = ST\boldsymbol{J} - k\nabla T \tag{3}$$

$$\boldsymbol{J} = -\sigma(\nabla V + S\nabla T) \tag{4}$$

where $\boldsymbol{q}$ is heat flux vector, $Q_e$ is Joule heat, $\boldsymbol{J}$ is current density vector, $\boldsymbol{E}$ is electric field vector, $S$ is Seebeck coefficient, $T$ is absolute temperature, $k$ is thermal conductivity, $V$ is voltage and $\sigma$ is electrical conductivity. Eq. 1 and Eq. 2 represent heat transfer in a thermoelectric material, where Joule heating is considered as the primary source of internal heat generation. Eq. 3 defines the total heat flux which combines the Peltier effect and traditional heat conduction dictated by Fourier's law. Eq. 4 describes the current density as driven by both electric potential and temperature gradients as a result of the Seebeck effect. In our TEC model, the thermal conductivity, Seebeck coefficient and electrical conductivity of holey silicon are set at 1 W/mK, 440 µV/K and $1.36 \times 10^4$ S/m, respectively, representing a 30% porosity, 20 nm neck size and highly doped P-type holey silicon thin film at elevated temperature[21,34,35]. The ambient temperature is set at 30 °C and the array boundary is set to be laterally conducted. Considering the target in transistor-scale thermal management, the size of one cooler unit is defined at $1 \times 1$ µm². Besides, to investigate the scaling effect and provide more flexibility, the TEC array can have arbitrary m rows and n columns ($1 \leq m, n \leq 10$) with random $N_{HS}$ assigned hotspots and $N_{TEC}$ assigned TECs ($0 \leq N_{HS}, N_{TEC} \leq \min[m \times n, 20]$). To simplify the problem while exaggerating the temperature difference, the assigned hotspot heat flux and assigned TEC voltage are pre-defined with 8 discrete intensities, ranging from $0.5 \times 10^{15}$ W/m³ to $4 \times 10^{15}$ W/m³ (an interval of $0.5 \times 10^{15}$ W/m³) and from 25mV to 200mV (an interval of 25mV), respectively. Fig. 1c demonstrates a specific example of using 2D steady-state FEM simulations for a $3 \times 3$ TEC array. From the simulations, a temperature map can be numerically derived given the values of m, n, $N_{HS}$, $N_{TEC}$ and their corresponding intensities.

### III. NEURAL NETWORK DEVELOPMENT

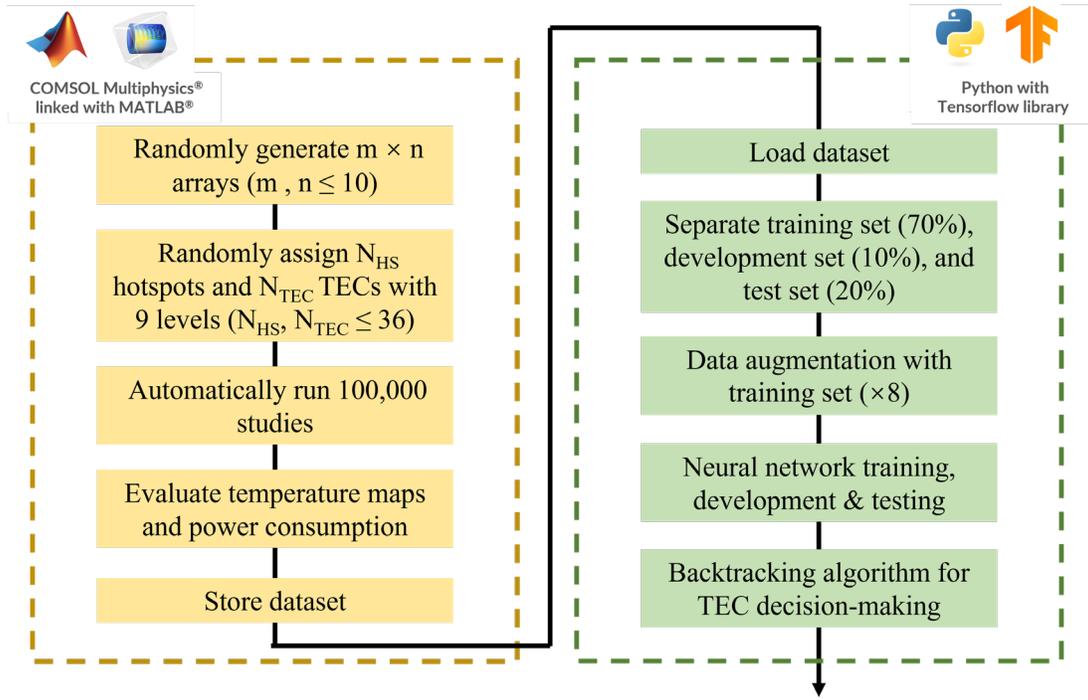

FIG. 2 Research workflow. (Left) Massive data generation based on autonomous FEM simulations and (right) Data postprocessing, including data splitting, data augmentation, neural network training and TEC optimization algorithm design.

Fig. 2 illustrates the research workflow which includes massive data generation and data postprocessing. Here, we utilize MATLAB R2021a to generate random hotspot and TEC inputs based on the given constraints, which will be sequentially fed to COMSOL 6.0 to conduct FEM simulations and evaluate temperature maps and power consumption as outputs. In MATLAB, the random values of rows (m) and columns (n) in the TEC array will be first defined. Later, two random m × n matrixes with values from 0 to 8 will be generated, representing the hotspot and TEC intensities. The maximum number of assigned hotspots ($N_{HS}$) and TECs ($N_{TEC}$) depend not only on the model constraints but also on the robustness of the training process. Here, we select min[m × n, 36] as the maximum number since it exceeds the original span ($0 \leq N_{HS}, N_{TEC} \leq$ min[m × n, 20]) for conducting a more robust training, meanwhile, it is not so large to weaken the independency of individual units. With the defined hotspot and TEC inputs, steady-state FEM simulations will be conducted, followed by the evaluation of temperature maps. Eventually, the temperature will be stored in an m × n output and the power consumption at 8 intensities will be in an 8 × 1 output. The autonomous program generates 100,000 random samples, whose total number considers the complexity of the model, the desired accuracy and the computation resources. After generating the original dataset, data splitting of training set (70%), development set (10 %) and test set (20%) is performed. The training set is further augmented with the transformation of flipping (horizontal and vertical), rotation (90°, 180° and 270°) and their combinations, resulting in a total number of training samples as 8 times as the original (560,000). This augmentation offers a cost-effective way to create new data, improve model's accuracy and robustness [36], and facilitate the learning of inherent symmetry.

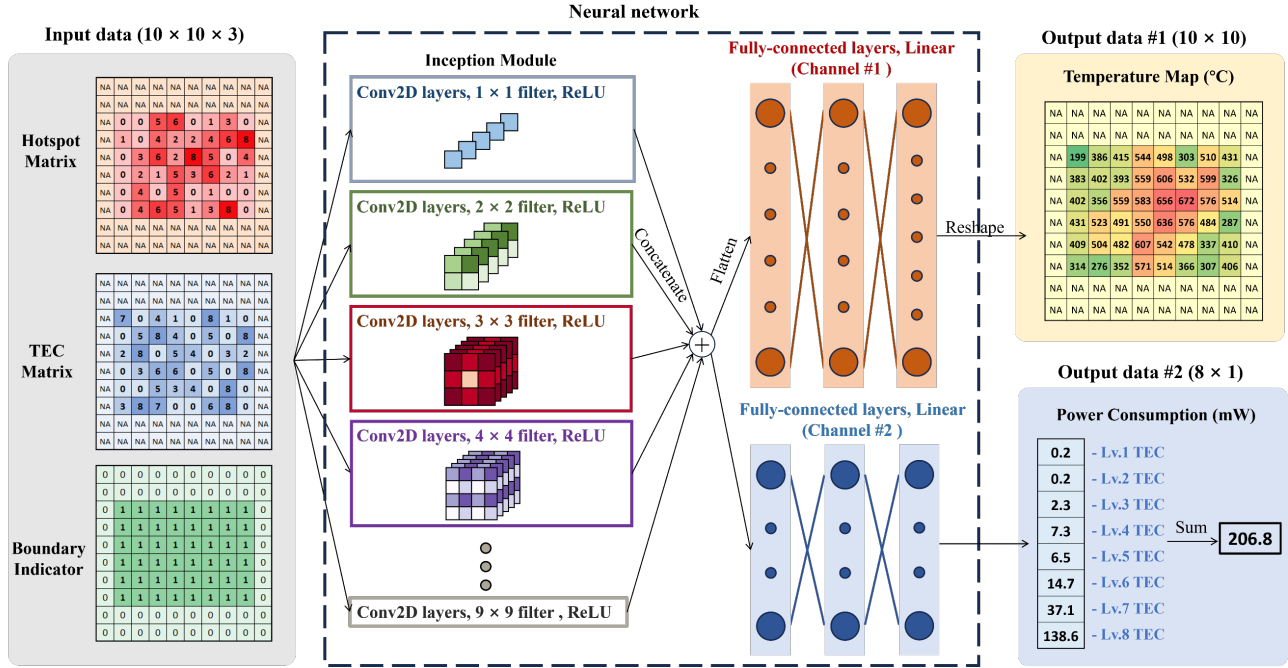

FIG. 3. CNN based on Inception module and multi-task learning. The input matrixes and output temperature matrix are padded into 10×10 arrays for uniform shapes. The output power matrix is an 8×1 array corresponding to 8 TEC intensities.

Due to the necessity to generate a 2D temperature map and its significant relevance to the spatial correlation among hotspots and TECs, a convolutional neural network (CNN) is chosen as the base model[37]. However, most CNN models in literature, such as LeNet-5[38], VGG16[39] and GoogLeNet (Inception)[40], are primarily designed for classification tasks. Therefore, modifications are necessary to adapt the CNN model for regression tasks with continuous output values. After conducting multiple tests and comparison, we have developed a CNN model based on the Inception[40,41] over other options (e.g., encoder-decoder [42,43] and residual network[44]) and employed a multi-task learning (MTL) approach[45,46] to comprehend the thermal-electrical physics underlying the system and attain accurate predictions for both temperature and power consumption, as depicted in Fig. 3. Specifically, the 10 × 10 × 3 input matrix consists of three channels, representing the hotspot heat flux, TEC voltage and boundary indicator. Note that the boundary indicator is introduced as an auxiliary input channel so that the model can learn its boundary effectively. For any m × n TEC array where m and n are less than 10, zero padding is applied to ensure uniform shapes for both input and temperature output. The Inception module incorporates nine pathways in parallel with one or more Conv2D layers in various kernel sizes. Each layer is then followed by a rectified linear unit (ReLu)[47] activation function to introduce non-linearity. Through many trials, it is observed that the network perform best when the layer counts are [1,3,4,4,1,1,1,1,1], corresponding to the pathways with kernel sizes from 1 × 1 to 9 × 9. More layers are found necessary for pathways with kernel sizes of 2 × 2, 3 × 3 and 4 × 4 since these kernels capture the most important spatial hierarchies in the system. Besides, the one-layer pathway with 1 × 1 kernel may play a role in boundary detection. Furthermore, other short pathways with large kernels seems crucial to perceive extensive spatial distribution[48]. The results obtained from multiple pathways are then concatenated and flattened before being processed by two separate series of fully-connected layers with linear activation functions for continuous output values. Finally, the first series generates 100 nodes, representing a 10 × 10 temperature map output, and the second series creates 8 nodes which correspond to the power consumption at 8 TEC intensities. During the training, a summation of two mean squared errors (MSE)[49,50] separately for temperature and power is defined as the loss function to adjust the weights for all layers through back-propagation, which can be written as:

$$L_{total} = MSE_T + MSE_P \qquad (5)$$

$$MSE_T = \frac{1}{M \times N} \sum_i^M \sum_j^N (Y_{i,j} - \hat{Y}_{i,j})^2 \qquad (6)$$

$$MSE_P = \frac{1}{N_L} \sum_k^{N_L} (Z_k - \hat{Z}_k)^2 \qquad (7)$$

where $L_{total}$ is the total loss, $MSE_T$ and $MSE_P$ are the mean squared errors of temperature and power consumption, respectively, $Y_{i,j}$ and $\hat{Y}_{i,j}$ are the ground-truth and predicted temperatures in the i$^{th}$ row and j$^{th}$ column, respectively, $M$ and $N$ are the maximum numbers of rows and columns, respectively (M = N =10), $Z_k$ and $\hat{Z}_k$ are the ground-truth and predicted power consumption values at the k$^{th}$ TEC intensity, respectively, and $N_L$ is the total number of non-zero TEC intensities ($N_L$ = 8). Fig. 4 demonstrate the MSE loss in 100 training epochs. With a learning rate of 0.001 and a batch size of 500 samples, the loss decreases rapidly through the first 20 epochs and stabilizes the third decimal place through the last 20 epochs. Table I summarizes the MSE loss for the training set, development set and test set.

While MSE is sensitive to large values and offers a smooth and differentiable function for the training, the mean absolute error (MAE)[51] provides a more interpretable evaluation of temperature prediction. The MAE loss is defined as:

$$MAE = \frac{1}{m \times n} \sum_i^m \sum_j^n |Y_{i,j} - \hat{Y}_{i,j}| \qquad (8)$$

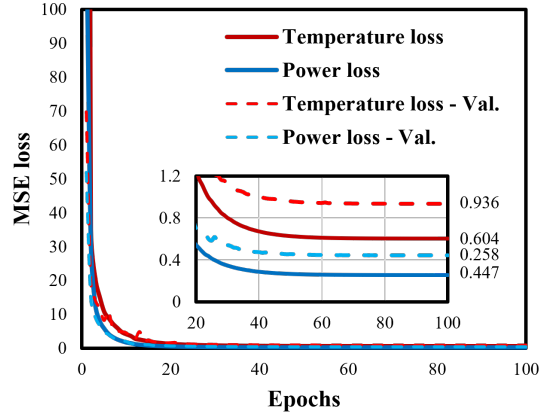

FIG. 4. MSE loss as a function of training epochs

TABLE I. Summary of MSE loss.

|  | Training Loss (70% data) | Validation Loss (10% data) | Test Loss (20% data) |
| --- | --- | --- | --- |
| **MSE - temperature** | 0.604 | 0.936 | 0.951 |
| **MSE - power** | 0.258 | 0.447 | 0.443 |

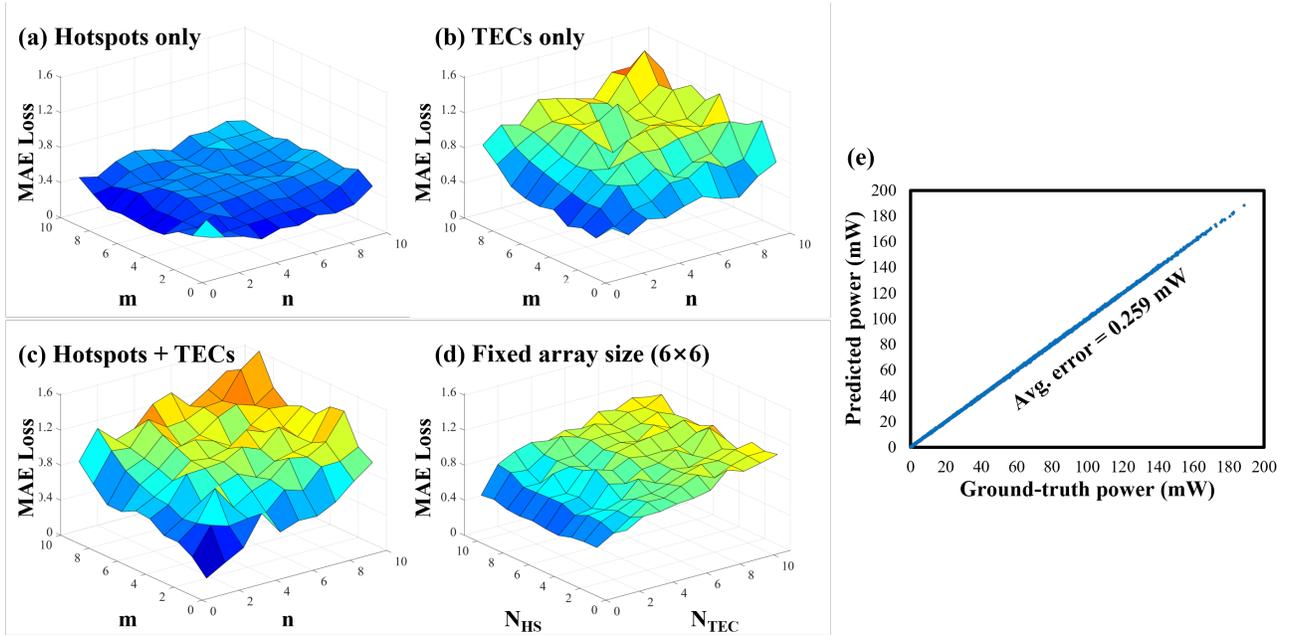

FIG. 5. Loss analysis of 4200 random samples. (a-c) MAE loss as a function of array dimensions ($1 \leq m, n \leq 10$) in 3000 samples. (a) Hotspots only. (b) TECs only. (c) Hotspots + TECs. (d) MAE loss as a function of hotspot and TEC counts ($1 \leq N_{HS}, N_{TEC} \leq 10$) in a $6 \times 6$ TEC array in 1200 samples. (e) Error of power consumption between ground-truth and predicted values for all samples.

where m and n represent the actual number of rows and columns before padding, respectively. Fig. 5a-e demonstrates the loss analysis with additional 4200 samples, where 10 random samples are taken into average for every combination of x-axis and y-axis arguments. As shown in Fig. 5a, a relatively small and uniform MAE loss occurs when no TEC is assigned. However, when any TEC is assigned, the MAE loss becomes generally large and will be impacted by the array rows and columns. This is because the implementation of thermoelectric effect significantly increases the model complexity compared to a pure heat transfer model. It should be noted that the number of assigned TECs does not significantly impact the MAE loss when more than one TEC is assigned, as shown in Fig. 5d. For all samples, the error of power consumption between the simulation and prediction values is plotted in Fig. 5e.

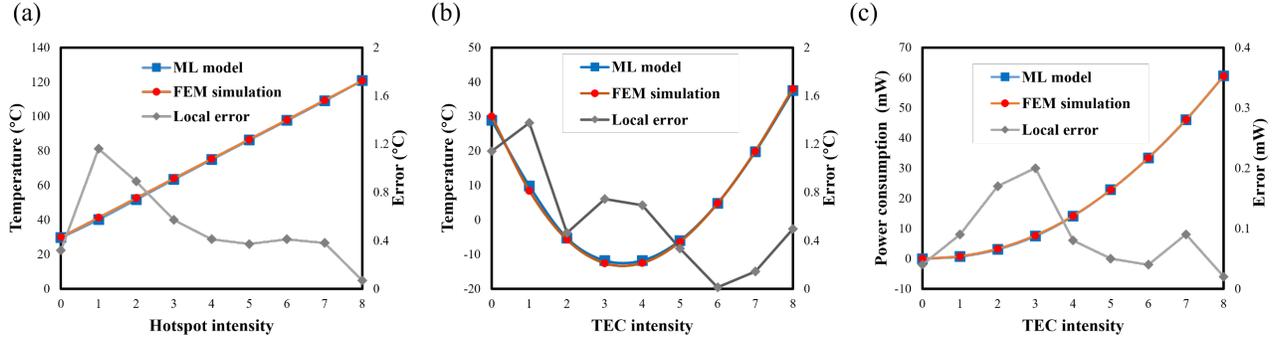

FIG. 6. Predictions of 1 × 1 array. (a) Temperature prediction for hotspot only. (b) Temperature prediction for TEC only. (c) Power consumption prediction for TEC only.

The predictions of a 1 × 1 TEC array (i.e., a single TEC) under hotspot only and TEC only scenarios are illustrated in Fig. 6. For temperature predictions, the MAE loss can be simply interpreted as the local error in a single TEC. The results show that the CNN model not only captures the proportional relationship between temperature and hotspot intensity, but it also successfully predicts the parabolic TEC cooling as a function of input voltage [50] due to coupled Joule and Peltier effect.

To demonstrate the multi-hotspot scenarios, Fig. 7 illustrates four prediction examples of a 6 × 6 TEC array with the following: (a) random hotspots, (b) random hotspots + TEC cooling, (c) clustered hotspots and (d) clustered hotspots + TEC cooling. In the first two scenarios, nine arbitrary hotspots are assigned with random intensities to represent a system incorporating different modules that experience various local heating conditions. In the last two scenarios, on the other hand, nine assigned hotspots are clustered within a 3 × 3 region with equal intensity to mimic a system consisting of similar components that undergo simultaneous workloads. It is observed that the orderly and clustered scenarios exhibit slightly higher MAE loss compared to the scattered and random scenarios due to the lower likelihood of generating well-organized data during data generation. Furthermore, for Fig. 7b and 7d, larger MAE loss is identified owing to the introduction of TEC cooling mechanism. Nevertheless, the key features of the TEC array can be safely captured: First of all, significant lateral heat redistribution can be observed from the predictions, where higher temperature occurs near the TEC-assigned regions compared to their hotspots-only counterparts. Secondly, the clustered TECs are predicted to have poorer effectiveness compared to the isolated TECs with the same intensity. This is because the adjacent TECs tend to generate active heat flow against each other, resulting in ineffective cooling. Lastly, local TEC cooling can be influenced by its corresponding hotspot conditions. A TEC is more likely to provide greater temperature reduction when its local hotspot has higher intensity. In summary, local TEC cooling will impact and be impacted by the surrounding TECs and hotspots. Therefore, achieving a global optimal solution is almost impossible by simply considering the local optimal TEC control.

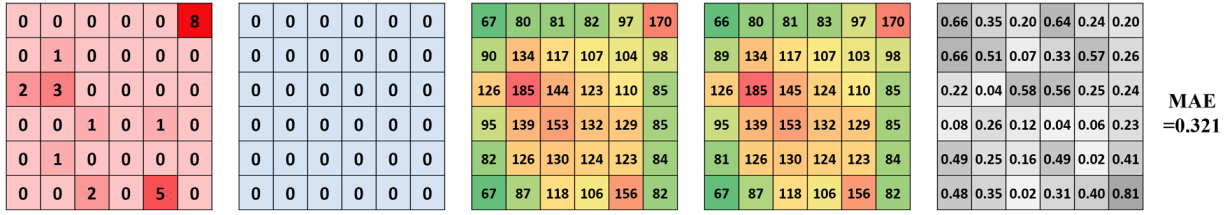
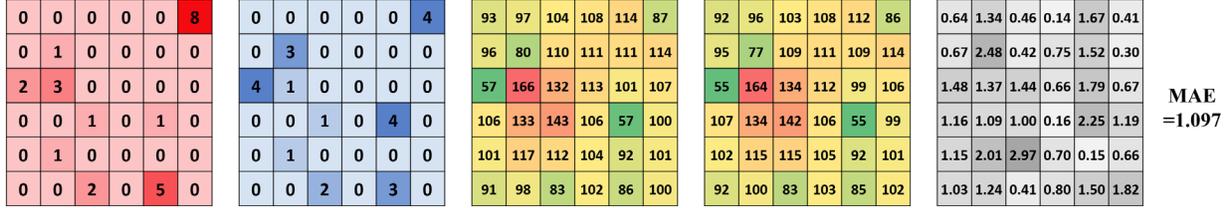
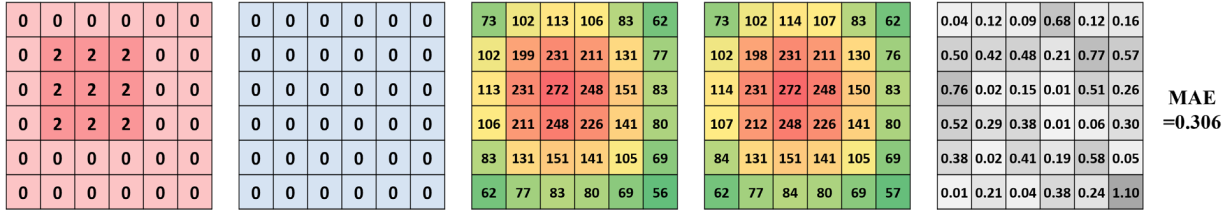
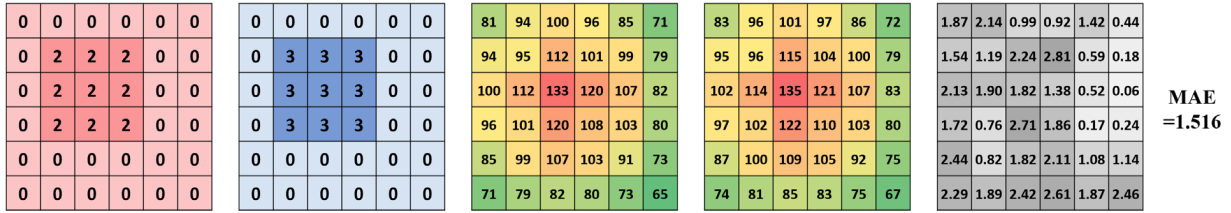

| Hotspot input | TEC input | FEM simulation | CNN prediction | Local error |

FIG. 7. Case studies of a 6 × 6 array for temperature predictions (unit: °C). (a) Arbitrary hotspots without TECs. (b) Arbitrary hotspots with TECs (not optimal). (c) Clustered hotspots without TECs. (d) Clustered hotspots with TECs (not optimal).

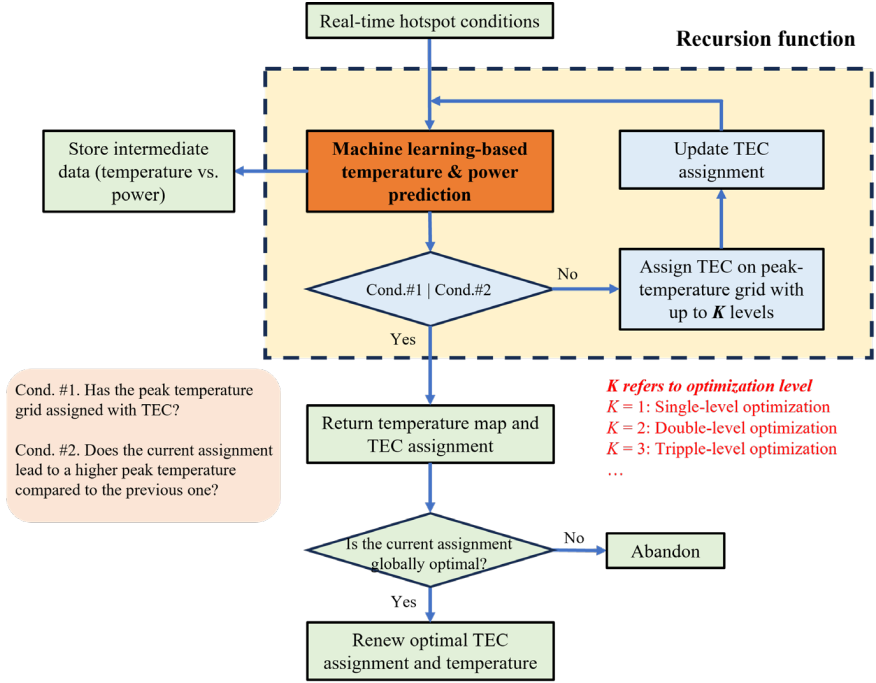

FIG. 8. Flowchart of backtracking-based TEC decision-making algorithm.

## IV. TEC OPTIMIZATION ALGORITHM

Due to the complex dependency among TEC array and multiple hotspots, using the traditional FEM-based techniques to enumerate all possible solutions seem difficult and even impossible. However, with the efficient optimization algorithm based on machine learning model, real-time global optimal solution can be feasible. In this study, we set our target to find the global optimal temperature (i.e., the smallest peak temperature) based on the existing hotspot conditions, while other possibilities, such as looking for the minimum TEC power/TEC counts for achieving acceptance temperature, can also be possible. Fig. 8 demonstrates the flowchart of the backtracking-based [52,53] TEC decision-making algorithm. This algorithm can compute the lowest peak temperature across the multi-hotspot system given the number of available TEC intensities (i.e. optimization level, $K$). The established CNN model serves as a function to efficiently evaluate the current status. To improve efficiency and reduce unnecessary iterations, two assumptions are made: first, the highest-temperature grid has the most priority to assign the TEC. Second, the next TEC assignment must lead to a lower peak temperature compared to the current one. Only when these two assumptions hold will the algorithm look for a deeper solution based on the existing ones. Fig. 9 demonstrates three cases of the 9 × 9 TEC array control using the developed algorithm: (a) random sparse hotspots, (b) random dense hotspots and (c) clustered hotspots. The peak temperatures of three samples (i.e., 348°C, 362°C, 349°C, respectively) experience substantial temperature reduction (i.e., dropped down by 177°C, 172°C and 176°C, respectively) after the single-level optimal TEC control. Here, we define cooling effectiveness = $(T_{peak,0} - T_{peak,opt})/(T_{peak,0} - 30°C)$, where $T_{peak,0}$ and $T_{peak,opt}$ are the original and optimal peak temperatures, respectively. At this point, the cooling effectiveness of three samples yields 55%, 52% and 55%, respectively. The total iteration counts (and times) are 26 (936ms), 44 (1584ms) and 47 (1692ms), respectively. A greater $K$ can lead to higher temperature reduction and cooling effectiveness at the expense of more computational cost. Additionally, all samples show a trend where the hotspot is moving towards the center as the optimization process evolves. This is because the TEC cooling inherently drives the system towards a more uniform temperature field, which manifests as the formation of a centralized hotspot with mitigated temperature gradient. It is worth noting that only small and moderate intensities of TEC (i.e., 1 to 5) are found in the optimal TEC assignments, while those large intensities of TEC (i.e., 6 to 8) may either cause too much penalty (i.e., temperature rise) to its neighboring or generate too much local Joule heat, which are abandoned by the optimization algorithm. This again demonstrates the fact that a local optimal TEC control may not suffice for the global optimal temperature. Interestingly, for clustered hotspots, optimal TEC assignments follow a staggered "checkerboard" pattern. This observation motivates a novel strategy for TEC array placement against uniform heat flux.

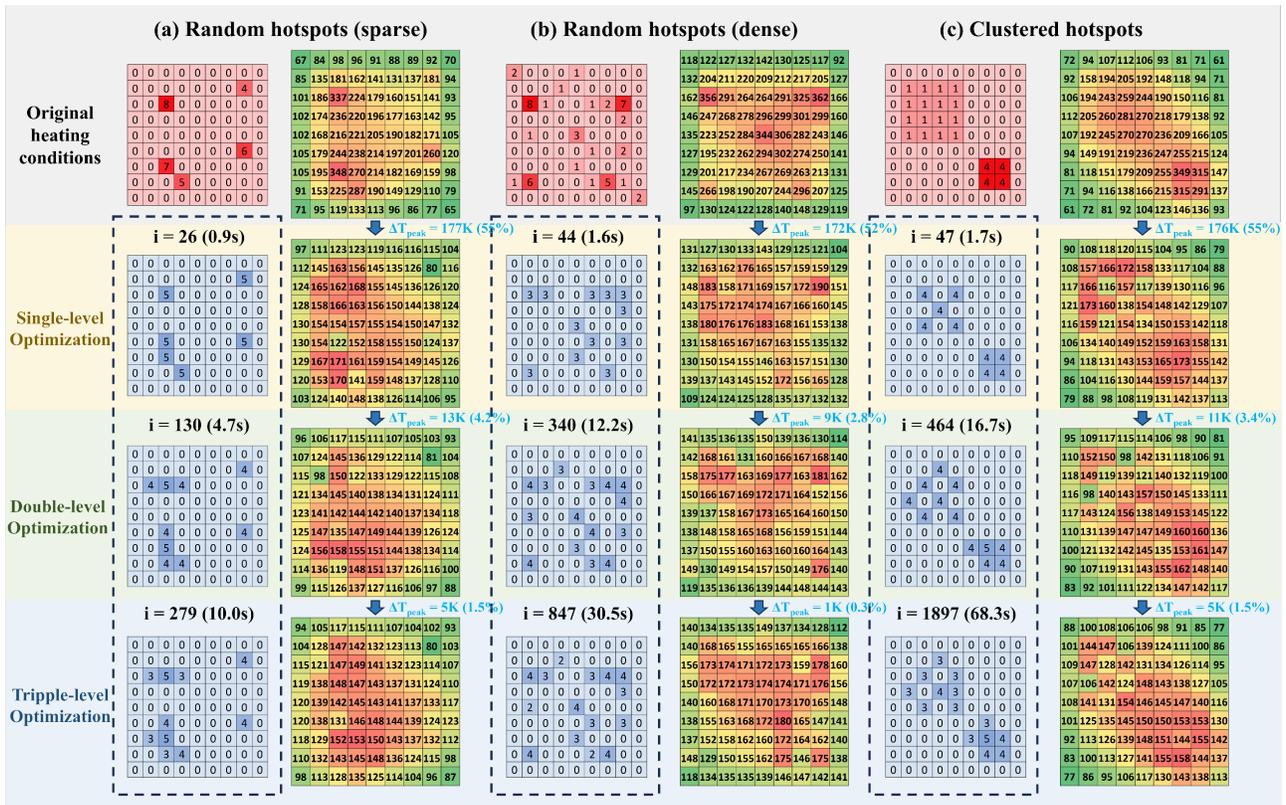

FIG. 9. Case studies of a 9 × 9 array for backtracking-based decision-making (unit: °C). The pink matrix represents the real-time hotspot conditions, the blue matrix represents the optimal TEC assignment based on the optimization level and the red-green matrix denotes the corresponding global optimal temperature map. Three scenarios are discussed: (a) random sparse hotspots, (b) random dense hotspots and (c) clustered hotspots.

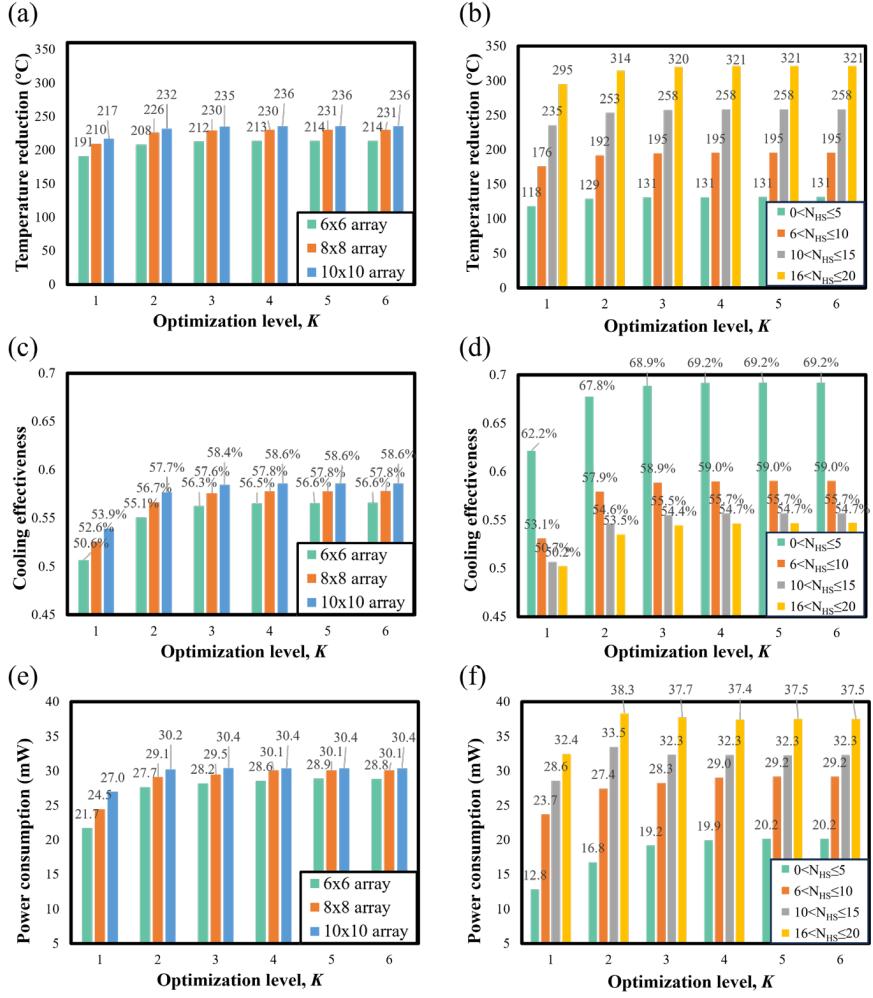

FIG. 10. As-achieved cooling performance and power consumption of 1800 random samples using machine learning-assisted TEC optimization algorithm. (a, b) Peak temperature reduction as a function of optimization level. (c, d) The corresponding cooling effectiveness. (e, f) The corresponding power consumption. The first column is varied by array dimensions and the second is by hotspot counts.

Fig. 10 evaluates the as-achieved cooling performance within 1800 random samples using the machine learning-assisted TEC optimization algorithm. A total of 600 samples are generated for each array size of 6 × 6, 8 × 8 and 10 × 10. Among these samples, there are 30 samples for every $N_{HS}$ ranging from 1 to 20. Here, the maximum optimization level is set at six, which allows up to six discrete intensities for the assigned TEC voltage. As a result, the average peak temperature reductions for the six levels are {206, 222, 226, 226, 226}°C, the average cooling effectiveness values are {52.3%, 56.5%, 57.4%, 57.6%, 57.6%, 57.6%}, and the corresponding power consumption yields {24.4, 29.0, 29.4, 29.7, 29.8, 29.8}mW, respectively. For $0 < K \leq 3$, an increase in $K$ provide a greater peak temperature reduction and higher cooling effectiveness at the expense of increased power consumption. However, for $3 < K \leq 6$, the cooling reaches a plateau. In general, larger arrays and smaller hotspot counts can result in more significant cooling due to more available TECs and larger space for heat redistribution.

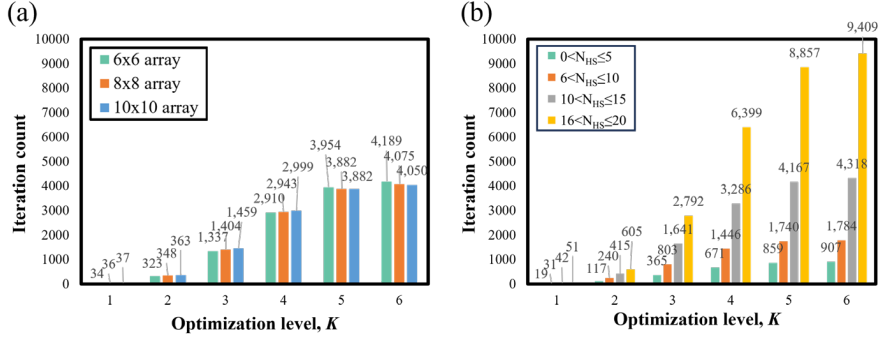

FIG. 11. Efficiency analysis of 1800 random samples using machine learning-assisted TEC optimization algorithm. (a) Iteration statistics of various array dimensions. (b) Iteration statistics of various hotspot counts.

Fig. 11 demonstrates the efficiency analysis based on the aforementioned 1800 samples using the machine learning-assisted TEC optimization algorithm. Here, the iteration count indicates the total number of predictions required to perform a single optimization, which reflects the computational cost. From level one to six, the average iteration counts are {36, 344, 1400, 2951, 3906, 4105}, respectively. The iteration count increases with the increased optimization level and becomes excessively large when $N_{HS}$ is large. Given the decreasing margin of cooling improvements, it is highly recommended to apply the TEC optimization algorithm with $K \leq 3$ in order to achieve balanced computational cost and TEC cooling performance.

To further investigate the efficiency of the machine learning-based TEC optimization algorithm, we record the single prediction time in Table. II for both FEM simulation and CNN model within 3000 samples as mentioned in Fig. 5, Following this, in Table. III, we summarize the running time for the optimization algorithm using the 1800 samples mentioned in Fig. 10 and Fig. 11. All FEM simulations are performed using COMSOL 6.0 with CPU computation on an AMD Ryzen 9 3950X processor (16-core, 3.5GHz) and 128GB memory, with a maximum of 1,582,714 degrees of freedom. The CNN model predictions are computed using GPU acceleration on a NVIDIA GeForce RTX 2080Ti (11GB) with a total of 124,157,612 parameters. Based on the statistical results, the average FEM simulation time for a single prediction is found to be 45s. Larger array sizes generally result in a longer computational time due to the increased degrees of freedom. Conversely, the CNN prediction demonstrates similar computational time through various input variables with an average time of only 42ms. A speed increase of over three orders of magnitude of is found when using the CNN model to conduct a single prediction compared to the traditional FEM methods. Furthermore, with the acceleration of the CNN model, the single-level, double-level and triple-level TEC optimization can be carried out within an average time of 1.5s, 14.5s and 58.8s, respectively, where the same amount of FEM computation will take about 26.8mins, 4.3hrs and 17.4hrs, respectively. The significant increases in speed paves the way for on-demand thermal management using realistic TEC systems. Future research will focus on the practical integration of TEC array into complex SoC systems, exploring ways to leverage machine-learning assisted TEC optimization algorithm to ensure efficient and reliable operation.

TABLE. II. Summary of single prediction time between simulation and CNN model

| m × n | Test samples | Maximum simulation time | Minimum simulation time | Average simulation time | Average CNN prediction time |
|---|---|---|---|---|---|
| [1,25] | 1530 | 39s | 4s | 20s | 42ms |
| [26,50] | 900 | 76s | 29s | 53s | 43ms |
| [51,75] | 390 | 108s | 69s | 89s | 42ms |
| [76,100] | 180 | 149s | 102s | 123s | 42ms |

TABLE. III. Summary of running time for machine learning-assisted TEC optimization

| Optimization levels | Test samples | Maximum time (iterations) | Minimum time (iterations) | Average time (iterations) |
|---|---|---|---|---|
| 1 | 1800 | 3.3s (78) | 0.3s (8) | 1.5s (36) |
| 2 | 1800 | 51s (1212) | 0.4s (9) | 14s (344) |
| 3 | 1800 | 282s (6702) | 0.4s (9) | 59s (1440) |
| 4 | 1800 | 815s (19403) | 0.4s (9) | 124s (2951) |
| 5 | 1800 | 1356s (32286) | 0.4s (9) | 164s (3906) |
| 6 | 1800 | 1545s (36780) | 0.4s (9) | 172s (4105) |

**V. CONCLUSIONS**

In this study, we present a novel machine learning-assisted TEC optimization algorithm aimed at achieving global optimal temperature control for on-demand multi-hotspot thermal management in microelectronic systems. Our findings demonstrate the ability of the machine learning-assisted algorithm to dynamically adapt to the evolving thermal landscape of microelectronic devices, efficiently offering optimal TEC control for managing the spatial and temporal variations of hotspots. The algorithm not only mitigates the computational burdens associated with the traditional FEM-based optimization techniques but also heralds a significant leap towards achieving the on-demand thermal management imperative for the sustainability and performance of advanced semiconductor devices.

ONFLICT OF INTEREST

The authors have no conflicts to disclose.

DATA AVAILABILITY

The data that support the findings of this study are available from the corresponding author upon reasonable request.